\documentstyle[12pt]{article}
\catcode`@=11
\newcount\@tempcntc
\def\@citex[#1]#2{\if@filesw\immediate\write\@auxout{\string\citation{#2}}\fi
  \@tempcnta\z@\@tempcntb\m@ne\def\@citea{}\@cite{\@for\@citeb:=#2\do
    {\@ifundefined
       {b@\@citeb}{\@citeo\@tempcntb\m@ne\@citea\def\@citea{,}{\bf ?}\@warning
       {Citation `\@citeb' on page \thepage \space undefined}}%
    {\setbox\z@\hbox{\global\@tempcntc0\csname b@\@citeb\endcsname\relax}%
     \ifnum\@tempcntc=\z@ \@citeo\@tempcntb\m@ne
       \@citea\def\@citea{,}\hbox{\csname b@\@citeb\endcsname}%
     \else
      \advance\@tempcntb\@ne
      \ifnum\@tempcntb=\@tempcntc
      \else\advance\@tempcntb\m@ne\@citeo
      \@tempcnta\@tempcntc\@tempcntb\@tempcntc\fi\fi}}\@citeo}{#1}}
\def\@citeo{\ifnum\@tempcnta>\@tempcntb\else\@citea\def\@citea{,}%
  \ifnum\@tempcnta=\@tempcntb\the\@tempcnta\else
   {\advance\@tempcnta\@ne\ifnum\@tempcnta=\@tempcntb \else \def\@citea{--}\fi
    \advance\@tempcnta\m@ne\the\@tempcnta\@citea\the\@tempcntb}\fi\fi}
\catcode`@=12
\def\be{\begin{equation}}
\def\ee{\end{equation}}
\def\bea{\begin{eqnarray}}
\def\eea{\end{eqnarray}}
\def\as{\alpha_s}
\def\CF{C_{\scriptscriptstyle F}}
\def\NC{N_{\scriptscriptstyle C}}
\def\eps{\varepsilon}
\def\ct{\cos\theta}
\def\cz{\chi_{\scriptscriptstyle Z}}
\def\MZ{M_{\scriptscriptstyle Z}}

\def\mbf#1{\mbox{\boldmath{$\displaystyle#1$}}}

\def\Li{\mbox{Li}_2}
\def\inted{\ln\left(\frac{1+v}{1-v}\right)}
\def\real{\mathop{\sl Re}\nolimits}
\def\imag{\mathop{\sl Im}\nolimits}
\font\promillefont=cmr8
\def\promille{\%\kern-.5pt\lower.5pt\hbox{\promillefont 0}}
\begin{document}
\thispagestyle{empty}
\begin{flushright}
MZ-TH/95-17 \\[-0.2cm]
hep-ph/9508399\\[-0.2cm]
August 1995 \\[-0.2cm]
\end{flushright}
\begin{center}

{\Large\bf Transverse Polarization of Top Quarks}\\[.3cm]
{\Large\bf Produced in {\boldmath$e^+e^-$}-Annihilation at
\boldmath{$O(\as)$}}\\[1.75cm]

{\large S.~Groote$^*$, J.G.~K\"orner\footnote{Supported in part
by the BMFT, FRG, under contract 06MZ566,\\
\hbox{\qquad}and by HUCAM, EU, under contract CHRX-CT94-0579}} \\[.4cm]
Institut f\"ur Physik, Johannes-Gutenberg-Universit\"at \\
Staudingerweg 7, D-55099 Mainz, Germany.

\end{center}
\vspace{1.5cm}
\centerline {\bf ABSTRACT}
\noindent
We present the results of an $O(\as)$ calculation of the transverse
polarization of top quarks produced in $e^+e^-$-annihilation. In a first
step we determine the transverse polarization of the top with regard to the
hadron plane spanned by the $(q,\overline{q},g)$ system. We then rotate the
transverse components of the polarization to the lepton plane spanned by
the $(q,e^+,e^-)$ system. After azimuthal averaging we determine the three
remaining inclusive transversely polarized structure functions. Together
with the one-loop and Born term contributions they determine the $\sin\theta$
and $\sin2\theta$ beam-quark polar angle dependence of the transverse
polarization. We present analytic and numerical results for the polarized
structure functions and the polar angle dependence of the transverse
polarization. We briefly comment on the transverse polarization of bottom
quarks produced in $e^+e^-$-annihilation.

\newpage
\noindent
The recent discovery of the top quark at Fermilab in $p\bar p$-collisions
provides the challenge and motivation to further investigate its production
and decay characteristics in other processes. A very convenient tool in this
regard is the proposed linear $e^+e^-$-collider that has sufficient energy
to produce top quark pairs. The produced top quarks in $e^+e^-$-annihilations
will be highly polarized. Furthermore they are so heavy that they decay
before hadronizing. Thus the measurement of the polarization components of
the produced top quarks is feasible through the study of spin-momentum
correlations in top quark decay~\cite{trans1}. In this note we will be
concerned with the transverse components of the top quark's polarization.
The Born term contributions to the top quark's transverse polarization have
been computed some time ago~\cite{trans2,trans3}. We shall present one-loop
$O(\as)$ corrections to these results. We mention that in particular the
transverse normal polarization has been widely discussed in the last few
years because it is a $T$-odd observable and thus has implications for the
possible observation of $CP$-violation in this process.
\bigskip\\\indent
For the three-body process
$(\gamma_V,Z)\rightarrow q(p_1)+\overline{q}(p_2)+g(p_3)$ we define a
polarized hadron tensor according to ($q=p_1+p_2+p_3$)
\be
H_{\mu\nu}(q,p_1,p_2,s)=\sum_{\overline{q},g\mbox{\scriptsize\ spins}}
  \langle q\,\overline{q}\,g|j_\mu|0\rangle\langle 0|j_\nu^\dagger|
  q\,\overline{q}\,g\rangle
\ee
Note that the spin sum does not include the quark's spin which one wants to
observe. The hadronic tensor $H_{\mu\nu}(q,p_1,p_2,s)$ can be decomposed
into a number of spin dependent and spin independent structure functions
which depend on $q^2$ and on the two energy variables $y=1-2p_1\!\cdot\!q/q^2$
and $z=1-2p_2\!\cdot\!q/q^2$. For our purposes it is convenient to work in
terms of helicity structure functions which we shall sometimes refer to as
rate functions.

In Table~1 we have listed a complete set of three-body helicity structure
functions both in terms of the helicity and the Cartesian components of the
hadron tensor. We have also listed the angular coefficients that multiply
the rate functions after contraction of the hadron and the lepton tensor.
The relative hadron-lepton orientation angles $\theta$ (polar) and $\chi$
(azimuthal) are defined in Fig.~1. In the following we shall refer to the
plane spanned by $(q,\overline{q},g)$ as the hadron plane and the plane
spanned by $(q,l^+,l^-)$ as the lepton plane. In this paper we restrict
our analysis to the case of unpolarized $e^+$ and $e^-$ beams. Longitudinal
beam polarization effects can easily be incorporated into the analysis since
they come in with the same angular dependence as written down in Table~1
(see e.g. \cite{trans2} and \cite{trans4}). Transverse beam polarization
effects introduce new angular dependencies which must be treated separately
\cite{trans4,trans5}.

There are in general nine independent components of the hadron tensor each
for the unpolarized case and for the three polarization directions. It is
quite apparent that the three-particle hadron tensor
$H_{\mu\nu}(q,p_1,p_2,s)$ posesses a very rich structure which can be
resolved in terms of the angular dependence given in Table~1.

Let us start by listing the tree-graph contribution to the three-body
hadron tensor. Here we limit our attention to the spin-dependent pieces.
The spin-independent pieces are given in~\cite{trans5}. For the
vector/vector ($VV$) and axial-vector/axial-vector ($AA$) contribution
we obtain (using the abbreviation $\xi=4m^2/q^2$)
\bea
H_{\mu\nu}^1&:=&\frac12(H_{\mu\nu}^{VV}+H_{\mu\nu}^{AA})\nonumber\\
  &=&\frac{2im}{y^2z^2q^4}\Bigg[\ q^2(2-\xi)y^2\eps(\mu\nu p_1s)
  -q^2(4yz-2\xi yz+2y^2z-2yz^2)\eps(\mu\nu p_2s)\nonumber\\&&
  +q^2(2y^2-\xi y^2-2yz+2\xi yz+\xi z^2)\eps(\mu\nu p_3s)\nonumber\\&&
  -4y^2(p_{1\mu}+p_{3\mu})\eps(\nu p_1p_2s)
  +4y^2(p_{1\nu}+p_{3\nu})\eps(\mu p_1p_2s)\nonumber\\&&
  +4y(yp_{1\mu}+zp_{2\mu}+yp_{3\mu})\eps(\nu p_2p_3s)
  -4y(yp_{1\nu}+zp_{2\nu}+yp_{3\nu})\eps(\mu p_2p_3s)
  \ \Bigg]\label{eqn1}\\
H_{\mu\nu}^2&:=&\frac12(H_{\mu\nu}^{VV}-H_{\mu\nu}^{AA})
  \ =\ \frac{2im}{y^2z^2q^4}\Bigg[\
  q^2(\xi y^2-2yz+\xi yz+\xi z^2+4yz^2)\eps(\mu\nu p_1s)\nonumber\\&&
  -q^2\xi z(y-z)\eps(\mu\nu p_2s)
  +q^2\xi y(y-z)\eps(\mu\nu p_3s)\nonumber\\&&
  -4yz(p_{1\mu}+p_{3\mu})\eps(\nu p_1p_2s)
  +4yz(p_{1\nu}+p_{3\nu})\eps(\mu p_1p_2s)\nonumber\\&&
  -4yz(p_{1\mu}-p_{2\mu}+p_{3\mu})\eps(\nu p_1p_3s)
  +4yz(p_{1\nu}-p_{2\nu}+p_{3\nu})\eps(\mu p_1p_3s)
  \ \Bigg].\label{eqn2}
\eea
Note that in general $H_{\mu\nu}^{VV}\ne H_{\mu\nu}^{AA}$ and thereby
$H_{\mu\nu}^2\ne 0$ in the massive quark case.

For the vector/axial-vector contributions we obtain
\bea
H_{\mu\nu}^3&:=&\frac i2(H_{\mu\nu}^{V\!A}-H_{\mu\nu}^{AV})\nonumber\\
  &=&\frac{2im}{y^2z^2q^4}\Bigg[
  -4yz(p_3s)(p_{1\mu}p_{2\nu}-p_{2\mu}p_{1\nu})
  +4yz((p_2s)+(p_3s))(p_{1\mu}p_{3\nu}-p_{3\mu}p_{1\nu})\nonumber\\&&
  -q^2(\xi y^2-4yz+2\xi yz+2y^2z+\xi z^2+4yz^2)
  (p_{1\mu}s_\nu-s_\mu p_{1\nu})\nonumber\\&&
  -2q^2yz^2(p_{2\mu}s_\nu-s_\mu p_{2\nu})
  -q^2(\xi y^2-2yz+\xi yz+2y^2z)(p_{3\mu}s_\nu-s_\mu p_{3\nu})
  \ \Bigg]\label{eqn3}\\ \nonumber
  \\
H_{\mu\nu}^4&:=&\frac12(H_{\mu\nu}^{V\!A}+H_{\mu\nu}^{AV})\nonumber\\
  &=&\frac{2m}{y^2z^2q^4}\Bigg[
  -q^2((\xi y^2-4yz+2\xi yz+2y^2z+\xi z^2+2yz^2)(p_2s)
  \nonumber\\&&\qquad\qquad
  +(-2y^2+\xi y^2-2yz+2\xi yz+4y^2z+\xi z^2)(p_3s))g_{\mu\nu}\nonumber\\&&
  -4y^2(p_3s)(p_{1\mu}p_{2\nu}+p_{2\mu}p_{1\nu})
  +8yz(p_3s)p_{2\mu}p_{2\nu}\nonumber\\&&
  -4y(z(p_2s)+y(p_3s))(p_{2\mu}p_{3\nu}+p_{3\mu}p_{2\nu})
  +2q^2y^2z(p_{1\mu}s_\nu+s_\mu p_{1\nu})\nonumber\\&&
  +q^2(\xi y^2-4yz+2\xi yz+4y^2z+\xi z^2+2yz^2)(p_{2\mu}s_\nu+s_\mu p_{2\nu})
  \nonumber\\&&
  +q^2(-2yz+\xi yz+2y^2z+\xi z^2)(p_{3\mu}s_\nu+s_\mu p_{3\nu})
  \ \Bigg].\label{eqn4}
\eea
The spin-dependent pieces of the structure functions $H_{\mu\nu}^1$,
$H_{\mu\nu}^2$ and $H_{\mu\nu}^3$ are antisymmetric in the Lorentz indices
$\mu$ and $\nu$, whereas the spin-dependent piece of $H_{\mu\nu}^4$ is
symmetric. Note that all the spin-dependent tree-graph contributions
$H_{\mu\nu}^i$ ($i=1,2,3,4$) are proportional to the quark mass. This means
that the transverse polarization vanishes in the mass zero limit. In the
case of the alignment polarization, however, the mass factor is cancelled
by the  denominator mass factor in the covariant polarization vector
$s_\mu^\ell$ and thus the alignment polarization survives in the mass zero
limit.

The three orthogonal polarization components of the quark are specified as
$s^\ell_\mu$ (alignment polarization along the momentum direction of the
quark), $s^\perp_\mu$ (transverse polarization in the hadron plane with
$s^\perp_\mu p_2^\mu\le 0$) and $s^N_\mu$ (transverse polarization normal
to the hadron plane with $(s^\perp_\mu,s^N_\mu,s^\ell_\mu)$ forming a
right-handed system in the quark's rest system). We then define polarized
helicity structure functions $H_\alpha^m$ ($m=\perp,N,\ell$; in the
following we shall mostly suppress the Lorentz index on $s^m_\mu$)
according to
\be\label{eqn5}
H_\alpha^{i,m}=H_\alpha^i(+s^m)-H_\alpha^i(-s^m),
\ee
where $\alpha=U,L,T,I,9,F,A,4,5$ label the nine independent components of
the hadron tensor (see Table~1). The unpolarized structure functions (not
explicitly shown here) are given by
\be
H_\alpha^i=H_\alpha^i(+s^m)+H_\alpha^i(-s^m)
\ee
and are independent of the choice of $s^m$. An inspection of the tree-level
expressions shows that not all of the polarized structure functions in
Table~1 are populated (seven each for $s^\ell$ and $s^\perp$
$(U,L,T,I,F,A,9)$ and five for $s^N$ $(4,5,A,F,9)$). The remaining
structure would be populated by absorbtive and/or $CP$-violating
contributions.

Our main interest in this paper is the transverse polarization of the quark
relative to the lepton plane after integration over the relative azimuthal
angle~$\chi$ of the lepton and hadron planes. This constitutes a more
inclusive polarization measure as compared to the full structure implied
by Eqs.~(\ref{eqn1}), (\ref{eqn2}), (\ref{eqn3}) and (\ref{eqn4}). It is
apparent from Fig.~1 that the hadron plane is rotated into the lepton plane
by the azimuthal angle $\chi$. The alignment polarization structure
function $H_\alpha^\ell$ is not affected by this rotation whereas the
transverse pieces are transformed according to
\bea
H_\alpha^{\perp'}&=&\cos\chi H_\alpha^\perp-\sin\chi H_\alpha^N
  \nonumber\\\label{eqn6}
H_\alpha^{N'}&=&\sin\chi H_\alpha^\perp+\cos\chi H_\alpha^N,
\eea
where the primed transverse polarization directions ${s^\perp}'$ and
${s^N}'$ now refer to the lepton plane. From the azimuthal
$\chi$-dependencies given in Table~1 and from Eq.~(\ref{eqn6}) one can
surmise that after the azimuthal integration
\begin{enumerate}
\item all transverse components $H_{U,L,T,F}^\perp$ and $H_{4,F}^N$ drop out,
\item there is a contribution {\em normal\/} to the lepton plane, i.e.
$\sigma^{N'}\ne 0$, coming from the imaginary part of the Breit-Wigner
resonance shape via $(\gamma,Z)$ interference. This contribution is of
order $O(\imag\cz(q^2)/\real\cz(q^2))$ and can thus safely be
neglected for top pair production. For example, in the threshold region of
top pair production, this transverse normal polarization effect is already
quite small since the factor $\imag\cz/\real\cz$ is approximately $0.1\%$
and decreases further with a $1/q^2$ power fall-off behaviour. We shall
nevertheless explicitly include this part in the following for the sake of
completeness and also because of the fact that this interference effect is
one of the sources of transverse normal polarization of $b$-quarks from
$Z$-decays. A further sizable contribution to $\sigma^{N'}$ comes from the
imaginary part of the one-loop graph which will be discussed later on.
\item $H_I^\perp$ and $H_5^N$ contribute with the weight factor
$\frac3{2\sqrt 2}\sin2\theta$, and $H_A^\perp$ and $H_9^N$ contribute with
the factor $\frac3{\sqrt 2}\sin\theta$ to the transverse polarization
{\em in\/} the lepton plane.
\end{enumerate}
For definiteness, we present our transverse polarization results in terms
of polarization cross sections. One has
\be
\frac{d\sigma^{\perp'}}{d\ct\,dy\,dz}
  =-\frac3{2\sqrt2}\sin2\theta\,g_{14}\frac{d\sigma_I^{4\perp'}}{dy\,dz}
  -\frac3{\sqrt2}\sin\theta\left(g_{41}\frac{d\sigma_A^{1\perp'}}{dy\,dz}
  +g_{42}\frac{d\sigma_A^{2\perp'}}{dy\,dz}\right),\label{eqn7}
\ee
\be
\frac{d\sigma^{N'}}{d\ct\,dy\,dz}
  =-\frac3{\sqrt2}\sin\theta\,g_{43}\frac{d\sigma_A^{3N'}}{dy\,dz},
  \label{eqn8}
\ee
where, in terms of the hadron-plane helicity structure functions
$H_\alpha^{i,m}$ defined in Eq.~(\ref{eqn5}), one has
\be\label{eqn9}
\frac{d\sigma_\alpha^{i,m'}}{dy\,dz}=\frac{\pi\alpha^2v}{3q^4}\Bigg\{
  \frac{q^2}{16\pi^2v}H_\alpha^{i,m'}\Bigg\}
\ee
with
\be
H_I^{4\perp'}=\frac12(H_I^{4\perp}+H_5^{4N})\qquad
H_A^{1,2\,\perp'}=\frac12(H_A^{1,2\,\perp}+H_9^{1,2\,N})\qquad
H_A^{3N'}=\frac12(H_A^{3N}-H_9^{3\perp}).
\ee
In Eq.~(\ref{eqn9}) we have split up the three-body phase space factor
into the two-body phase space factor $\pi\alpha^2v/3q^4$ and the relative
two-body/three-body phase space factor $q^2/16\pi^2v$ in order to facilitate
the comparison with the respective two-body loop and Born term
contributions.

The respective polarization cross sections represent components of the
unnormalized polarization vector. The corresponding components of the
normalized polarization vector are then obtained by dividing by the
unpolarized cross section given by
\bea
\frac{d\sigma}{d\ct\,dy\,dz}
  &=&\frac38(1+\cos^2\theta)\left(g_{11}\frac{d\sigma_U^1}{dy\,dz}
  +g_{12}\frac{d\sigma_U^2}{dy\,dz}\right)\nonumber\\&&
  +\frac34\sin^2\theta\left(g_{11}\frac{d\sigma_L^1}{dy\,dz}
  +g_{12}\frac{d\sigma_L^2}{dy\,dz}\right)
  +\frac34\cos\theta g_{44}\frac{d\sigma_F^4}{dy\,dz},
  \label{eqn10}
\eea
where, in analogy with Eq.~(\ref{eqn9}), one has
\be
\frac{d\sigma_\alpha^i}{dy\,dz}=\frac{\pi\alpha^2v}{3q^4}\Bigg\{
  \frac{q^2}{16\pi^2v}H_\alpha^i\Bigg\}.
\ee
The unpolarized cross section results are taken from \cite{trans5,trans6}
and will not be listed explicitly in this paper.

\newpage

In Eqs.~(\ref{eqn7}), (\ref{eqn8}) and (\ref{eqn10}), we have perused
the electro-weak coupling parameters
\bea
g_{11}&=&Q_f^2-2Q_fv_ev_f\real\cz+(v_e^2+a_e^2)(v_f^2+a_f^2)|\cz|^2,
  \nonumber\\
g_{12}&=&Q_f^2-2Q_fv_ev_f\real\cz+(v_e^2+a_e^2)(v_f^2-a_f^2)|\cz|^2,
\nonumber\\
g_{13}&=&-2Q_fv_ea_f\imag\cz,\nonumber\\
g_{14}&=&2Q_fv_ea_f\real\cz-2(v_e^2+a_e^2)v_fa_f|\cz|^2,\nonumber\\
g_{41}&=&2Q_fa_ev_f\real\cz-2v_ea_e(v_f^2+a_f^2)|\cz|^2,\\
g_{42}&=&2Q_fa_ev_f\real\cz-2v_ea_e(v_f^2-a_f^2)|\cz|^2,\nonumber\\
g_{43}&=&2Q_fa_ea_f\imag\cz,\nonumber\\
g_{44}&=&-2Q_fa_ea_f\real\cz+4v_ea_ev_fa_f|\cz|^2\nonumber
\eea
with
\be
\cz(q^2)=\frac{gM_Z^2q^2}{q^2-M_Z^2+iM_Z\Gamma_Z},\quad
g=\frac{G_F}{8\sqrt 2\pi\alpha}\approx 4.49\cdot 10^{-5}
\mbox{\rm GeV}^{-2}.
\ee
The Standard Model values of the electro-weak coupling constants are given
by
\bea
v_e&=&-1+4\sin^2\theta_W,\quad a_e\ =\ -1\quad
  \hbox{\rm for leptons},\nonumber\\
v_f&=&1-\frac83\sin^2\theta_W,\quad a_f\ =\ 1\quad
  \hbox{\rm for up-type quarks ($Q_f=\frac23$), and}\\
v_f&=&-1+\frac43\sin^2\theta_W,\quad a_f\ =\ -1\quad
  \hbox{\rm for down-type quarks ($Q_f=-\frac13$)}.\nonumber
\eea

It is quite instructive to perform the above inclusive analysis more
directly using the covariant decomposition of the relevant inclusive hadron
tensor. One has
\bea
\lefteqn{\frac{q^2}{16\pi^2v}\int_0^{2\pi}\frac{d\chi}{2\pi}
  \int_{z_-(y)}^{z_+(y)}dz\,H_{\mu\nu}(q,p_1,p_2,s)\ =}\nonumber\\
  &\!=\!&\mbf{-\hat g_{\mu\nu}H_1^{pc}+\hat p_{1\mu}\hat p_{1\nu}H_2^{pc}
  +i\eps(\mu\nu p_1q)H_3^{pv}}
  +q_\mu q_\nu H_4^{pc}+(q_\mu p_{1\nu}+q_\nu p_{1\mu})H_5^{pc}\nonumber\\&&
  +(q\!\cdot\!s)\!\left[\mbf{-\hat g_{\mu\nu}G_1^{pv}
  +\hat p_{1\mu}\hat p_{1\nu}G_2^{pv}+i\eps(\mu\nu p_1q)G_3^{pc}}
  +q_\mu q_\nu G_4^{pv}+(q_\mu p_{1\nu}+q_\nu p_{1\mu})G_5^{pv}\right]
  \nonumber\\&& \mbf{+(s_\mu\hat p_{1\nu}+s_\nu\hat p_{1\mu})G_6^{pv}}
  +(s_\mu q_\nu+s_\nu q_\mu)G_7^{pv}
  \mbf{+i\eps(\mu\nu qs)G_8^{pc}+i\eps(\mu\nu\hat p_1s)G_9^{pc}}\nonumber\\&&
  \mbf{+i(s_\mu\hat p_{1\nu}-s_\nu\hat p_{1\mu})G_{10}^{pv}}
  +i(s_\mu q_\nu-s_\nu q_\mu)G_{11}^{pv}\nonumber\\&&
  \mbf{+(\hat p_{1\mu}\eps(\nu qp_1s)+\hat p_{1\nu}\eps(\mu qp_1s))
  G_{12}^{pc}}+(q_\mu\eps(\nu qp_1s)+q_\nu\eps(\mu qp_1s))G_{13}^{pc}.
  \label{eqn11}
\eea
The spin vector $s_\mu$ now refers to any {\em fixed\/} space coordinate
system. In particular, the fixed system should not make reference to the
hadron plane since one is integrating over the relative lepton-hadron
azimuth in the inclusive measure Eq.~(\ref{eqn11}). For our purposes it is
most convenient to choose the lepton system $(x',y',z'(=z))$ as the
reference system for the spin vector with the electron momentum pointing in
the negative $x'$-direction, i.e. Eq.~(\ref{eqn11}) should be read
with the replacement $s_\mu\rightarrow s_\mu'$. There are altogether five
spin-independent and 13 spin-dependent structure functions. Of these only
the eleven boldfaced contribute to $e^+e^-$-annihilation for zero lepton mass.
As mentioned before, the remaining seven structure functions do not vanish
by any means but are of no interest in this reaction since they cannot be
probed in the zero lepton mass case. It is for this reason that we have
written the covariants in Eq.~(\ref{eqn11}) in terms of the four-transverse
objects $\hat g_{\mu\nu}=g_{\mu\nu}-q_\mu q_\nu/q^2$ and
$\hat p_{1\mu}=p_{1\mu}-(p_1\!\cdot q)q_\mu/q^2$ so that the invariants
separate into the two mentioned categories. Also the invariant structure
functions $G_8$ and $G_9$ have been defined such that only $G_8$ is
contributed to by {\em transversely\/} polarized quarks. For quick
identification, the invariants carry a superscript ($pc$: parity conserving;
$pv$: parity violating) according to whether they are fed by the parity
conserving $VV$, $AA$ or by the parity violating $V\!A$ products of the
hadron currents. This superscript will be dropped in the following. Let us
finally mention that $G_{10}$, $G_{11}$, $G_{12}$ and $G_{13}$ are so-called
$T$-odd structure functions. The tree-graph only contributes to the $T$-odd
structure functions $G_{10}$ and $G_{11}$.

The transverse polarization in the lepton plane is given in terms of the
two structure functions $G_6$ and $G_8$. These can easily be projected out
by contracting Eq.~(\ref{eqn11})
(with $s_\mu\rightarrow s_\mu'$) on the r.h.s. with
$({s_\mu^\perp}'\hat p_{1\nu}+{s_\nu^\perp}'\hat p_{1\mu})$ and
$\eps(\mu\nu q{s^\perp}')$, respectively. On the l.h.s. of Eq.~(\ref{eqn11})
the contractions can be done directly on the integrand since integration
and contraction commute. In this way we obtain
\bea
-\sqrt2 p_{1z}G_6^4&=&\frac{q^2}{16\pi^2v}\int\frac{d\chi}{2\pi}
  \int dz\frac12(H_I^{4\perp}+H_5^{4N}),\\
\sqrt2 \sqrt{q^2}G_8^{1,2}&=&\frac{q^2}{16\pi^2v}\int\frac{d\chi}{2\pi}
  \int dz\frac12(H_A^{1,2\,\perp}+H_9^{1,2\,N}).
\eea
The transverse normal structure function $G_{10}$ is projected out by
contraction with $({s_\mu^N}'\hat p_{1\nu}-{s_\nu^N}'\hat p_{1\mu})$. One
obtains
\be
\sqrt2p_{1z}G_{10}^3=\frac{q^2}{16\pi^2v}\int\frac{d\chi}{2\pi}
  \int dz\frac12(H_A^{3N}-H_9^{3\perp}).
\ee

Let us now present our results for the integrated three-body tree-graph
polarized rate functions in terms of the rate functions
\be
\hat H_\alpha^{i,m'}(tree)=\frac{q^2}{16\pi^2v}\int dy\,dz\,
  H_\alpha^{i,m'}(y,z),
\ee
where the hat symbol on the structure function has been added to remind
one-self that one is dealing with an integrated three-body structure
function. The complete $O(\as)$ contribution is then given by adding in
the $O(\as)$ one-loop contributions (taken from~\cite{trans6}). One has
\be\label{eqn12}
\hat H_\alpha^{i,m'}(\as)=\hat H_\alpha^{i,m'}(tree)+H_\alpha^{i,m'}(loop).
\ee
Note that the IR-singularity contained in the integrated tree-graph
contribution exactly cancels the IR-singularity in the one-loop
contribution such that the sum of the two terms in Eq.~(\ref{eqn12}) is
IR-finite. We have regularized the IR-singularity by introducing a (small)
gluon mass. The gluon mass regulator is of no concern when calculating the
tree-graph matrix elements Eqs.~(\ref{eqn1}--\ref{eqn4}) but only serves
to slightly deform the phase-space away from the IR-singularity. One has
\medskip\\
{\it $O(\as)$ real part:}
\bea
\hat H_I^{4\perp'}(\as)
  &=&\frac12(\hat H_I^{4\perp}(\as)+\hat H_5^{4N}(\as))\nonumber\\
  &=&-\frac{\as}\pi\NC\CF\frac{m\sqrt{q^2}}{2\sqrt2v}\Big[
  (1-\sqrt\xi)(24-7\sqrt\xi+\frac32\xi)+4v^2t_{11}\nonumber\\&&
  +4v^2t_{10}-2(2-\xi)v(t_7-t_8)+(8+\frac72\xi)t_6\nonumber\\&&
  -(21+2\xi)vt_3-\frac18(4-\xi)(10+3\xi)(t_1-t_2)\Big]\label{eqn13}\\
\hat H_A^{1\perp'}(\as)
  &=&\frac12(\hat H_A^{1\perp}(\as)+\hat H_9^{1N}(\as))\nonumber\\
  &=&-\frac{\as}\pi\NC\CF\frac{m\sqrt{q^2}}{4\sqrt2v}\Big[
  (8-3\xi)v+16vt_{12}+8vt_{10}\nonumber\\&&
  +4(2-\xi)(t_8-t_9)-(8+\xi)t_5\nonumber\\&&
  +(1-\sqrt\xi)(8-2\sqrt\xi-\xi)t_4-(36-19\xi+\frac32\xi^2)t_3\Big]\\
\hat H_A^{2\perp'}(\as)
  &=&\frac12(\hat H_A^{2\perp}(\as)+\hat H_9^{2N}(\as))\nonumber\\
  &=&\frac{\as}\pi\NC\CF\frac{m\sqrt{q^2}}{4\sqrt2v}\Big[
  (20-3\xi)v-16vt_{12}-8vt_{10}\nonumber\\&&
  -4(2-\xi)(t_8-t_9)+\xi t_5-\xi(1-\sqrt\xi)t_4+(16-7\xi-\frac32\xi^2)t_3
  \Big]\\
\hat H_A^{3N'}(\as)
  &=&\frac12(\hat H_A^{3N}(\as)-\hat H_9^{3\perp}(\as))\nonumber\\
  &=&\frac{\as}\pi\NC\CF\frac{m\sqrt{q^2}}{2\sqrt2v}\Big[
  (1-\sqrt\xi)(10-3\sqrt\xi+\frac32\xi)-4v^2t_{11}\nonumber\\&&
  -4v^2t_{10}+2(2-\xi)v(t_7-t_8)-(4-\frac{13}2\xi)t_6\nonumber\\&&
  +(1-6\xi)vt_3-(3-\frac14\xi-\frac38\xi^2)(t_1-t_2)\Big]\label{eqn14}
\eea
Closed form expressions for the integrals $t_i$ ($i=1,\ldots,12$) appearing
in Eqs.~(\ref{eqn13})--(\ref{eqn14}) can be found in the Appendix.
\goodbreak
What remains is to write down the corresponding Born term expressions. The
easiest way to obtain these is to directly read off the relevant covariant
structure functions $G_j^i$ from the two-body hadron tensor $H_{\mu\nu}^i$.
One has
\medskip\\
{\it Born terms:}
\bea
-\sqrt2 p_{1z}G_6^4&=&H_I^{4\perp'}
  \ =\ \frac12(H_I^{4\perp}+H_5^{4N})\ =\ -\sqrt2\NC mv\sqrt{q^2}\\
\sqrt2 \sqrt{q^2}G_8^{1,2}&=&H_A^{1,2\,\perp'}
  \ =\ \frac12(H_A^{1,2\,\perp}+H_9^{1,2\,N})\ =\ \sqrt2\NC m\sqrt{q^2}\\
\sqrt2 p_{1z}G_{10}^3&=&H_A^{3N'}
  \ =\ \frac12(H_A^{3N}-H_9^{3\perp})\ =\ \sqrt2\NC mv\sqrt{q^2}
\eea
As mentioned before, the contribution of $H_A^{3N'}$ can be neglected for
$t\bar t$-production as the $t\bar t$-threshold is quite far away from the
$Z$-pole.

The last missing piece of information is the $O(\as)$ contribution of the
imaginary part of the one-loop contribution to
$\frac12(H_I^{iN}-H_5^{i\perp})\propto G_{12}^i$ ($i=1,2$),
$\frac12(H_A^{4\perp}-H_9^{4N})\propto G_{10}^4$ and
$\frac12(H_I^{3\perp}+H_5^{3N})\propto G_6^3$. These can be read off from
the one-loop result given e.g. in~\cite{trans6}.\\
{\it $O(\as)$ imaginary part:}
\bea
-\sqrt2 p_{1z}G_6^3\ =\ \hat H_I^{3\perp'}(\as)
  &=&\frac12(\hat H_I^{3\perp}(\as)+\hat H_5^{3N}(\as))\nonumber\\
  &=&-\frac{\as}\pi\NC\CF\frac{m\sqrt{q^2}}{2\sqrt2}\pi(1+\xi)\\
-2\sqrt2 mp_{1z}^2G_{12}^{1,2}\ =\ \hat H_I^{1,2\,N'}(\as)
  &=&\frac12(\hat H_I^{1,2\,N}(\as)-\hat H_5^{1,2\,\perp}(\as))\nonumber\\
  &=&-\frac{\as}\pi\NC\CF\frac{m\sqrt{q^2}}{2\sqrt2}\pi v\label{eqn15}\\
\sqrt2 p_{1z}G_{10}^4\ =\ \hat H_A^{4N'}(\as)
  &=&\frac12(\hat H_A^{4N}(\as)-\hat H_9^{4\perp}(\as))\nonumber\\
  &=&\frac{\as}\pi\NC\CF\frac{m\sqrt{q^2}}{2\sqrt2}\pi(1+\xi).
\eea
The imaginary parts of the one-loop contribution for $\hat H_I^{1,2\,N'}$
and $\hat H_A^{4N'}$ agree with the results given in~\cite{trans2}. The
contribution to the transverse perpendicular structure function
$\hat H_I^{3\perp'}$ has not been considered in~\cite{trans2}. Of course,
in case of $t\bar t$-production, this contribution can again safely be
neglected. The angular factors entering the two-body polarization cross
sections can be read off from Table~1. One obtains
\be
\frac{d\sigma^{\perp'}}{d\ct}
  =\frac{\pi\alpha^2v}{3q^4}\Bigg\{-\frac3{2\sqrt2}\sin2\theta\,g_{13}
  \hat H_I^{3\perp'}\Bigg\}
\ee
\be
\frac{d\sigma^{N'}}{d\ct}
  =\frac{\pi\alpha^2v}{3q^4}\Bigg\{-\frac3{2\sqrt2}\sin2\theta
  (g_{11}\hat H_I^{1N'}+g_{12}\hat H_I^{2N'})
  -\frac3{\sqrt2}\sin\theta\,g_{44}\hat H_A^{4N'}\Bigg\}.
\ee

We are now in the position to determine the polar angle dependence of the
two transverse polarizations by adding up the $O(\as^0)$ Born term
contribution and the $O(\as)$ loop and tree contributions according to the
ratio expressions
\bea
P^{\perp'}(\ct)&=&\frac{d\sigma^{\perp'}(Born+loop+tree)/d\ct}%
  {d\sigma(Born+loop+tree)/d\ct}\\
P^{N'}(\ct)&=&\frac{d\sigma^{N'}(Born+loop+tree)/d\ct}%
  {d\sigma(Born+loop+tree)/d\ct}.
\eea
The polarized expressions constitute mean polarizations over the full
$(y,z)$-Dalitz plot region.

In Fig.~2a we show our results for the $O(\as^0)+O(\as)$ perpendicular
polarization of the top quark for three repesentative c.m. energies, where
the lowest energy $\sqrt{q^2}=360$~GeV is chosen to lie far enough above
the nominal threshold value of $\sqrt{q^2}=348$~GeV for a perturbative
calculation to make sense. The perpendicular polarization is positive and
large. It decreases with increasing energy because of the aforementioned
$m/\sqrt{q^2}$ dependence of the transverse polarization. We also show the
Born term results (dotted lines). It is apparent that the $O(\as)$
corrections to the Born term result are negative and small. At
$\sqrt{q^2}=1000$~GeV they can amount up to $\simeq 10\%$ depending on the
value of the polar angle~$\theta$. Close to threshold the radiative
corrections have become so small that they are hardly visible on the scale
of the figure. The perpendicular polarization is positive over the whole
angular range indicating that the $\sin\theta$ contribution overwhelms the
$\sin2\theta$ contribution, in particular close to threshold.

In Fig.~2b we show the polar angle dependence of the normal polarization,
again for the three energies $\sqrt{q^2}=360$~GeV, 500~GeV and 1000~GeV.
The polarization peaks towards the larger $\theta$-values, where the electron
and the top quark are in different hemispheres. The normal polarization
is mainly an $O(\as)$ effect coming from the imaginary part of the $O(\as)$
vertex correction. The corresponding $O(\as^0)$ Born term contributions are
not drawn since they are so small that they cannot be discerned from the
line of the abscissa. Again the $\sin\theta$ contribution dominates over the
$\sin2\theta$ contribution because of the presence of the threshold power
$v=\sqrt{1-\xi}$ in $\hat H_I^{1,2\,N'}$ (see Eq.~(\ref{eqn15})). The
linear $m/\sqrt{q^2}$ power behaviour of the transverse polatization is
clearly evident in Fig.~2b in as much as the polarization decreases with
increasing energy over most of the $\ct$-range.

In Fig.~3 we show the transverse polarization of the bottom quark on the
$Z$-peak. The perpendicular polarization (Fig.~3a) is small but still
sizeable which shows that one cannot always neglect bottom quark mass
effects. Also the radiative corrections are quite large in this case. In
the case of the bottom quark the $\sin2\theta$ term is the dominating term
in the polar angle distribution. In fact, when averaged over $\ct$, the
perpendicular polarization can be seen to be quite small
($\langle P^{\perp'}\rangle\simeq 0.61\%$).

The transverse normal polarization of bottom quarks in $Z\rightarrow b\bar b$
is shown in Fig.~3b. As it turns out both contributions from the imaginary
part of the loop and the Breit-Wigner interference term are of almost equal
importance with a slight dominance of the Breit-Wigner contribution
proportional to $\sin\theta$. We have taken a bottom quark mass of
$m_b=4.83$~GeV~\cite{trans7} for the curves in Fig.~3. If one uses a
running bottom quark mass $\overline{m_b}(\MZ)=2.69$~GeV as in our previous
works~\cite{trans6,trans8}, the transverse polarization is reduced by a
factor of approximately $m_b/\overline{m_b}(\MZ)\simeq 1.8$.
\vspace{1cm}\\
{\bf Acknowledgement: }We would like to thank K.G.\ Chetyrkin, K.\ Melnikov
and M.M.\ Tung for helpful discussions.
\newpage
\section*{Appendix}
\setcounter{equation}{0}
\def\theequation{A\arabic{equation}}
It is convenient to define the mass dependent variables $a:=2+\sqrt\xi$,
$b:=2-\sqrt\xi$ and $w:=\sqrt{(1-\sqrt\xi)/(1+\sqrt\xi)}$. The integrals
$t_1,\ldots,t_{12}$ appearing in Eqs.~(\ref{eqn13})--(\ref{eqn14}) are then
given by
\bea
t_1&:=&\ln\left(\frac{2\xi\sqrt\xi}{b^2(1+\sqrt\xi)}\right),\quad
t_2\ :=\ \ln\left(\frac{2\sqrt\xi}{1+\sqrt\xi}\right)\quad
  \Rightarrow\quad t_1-t_2\ =\ \ln\left(\frac\xi{b^2}\right)\\
t_3&:=&\inted\\
t_4&:=&\Li(w)-\Li(-w)+\Li(\frac abw)-\Li(-\frac abw)\\
t_5&:=&\frac12\ln\left(\frac{\sqrt\xi(2+\sqrt\xi)}{4(1+\sqrt\xi)}\right)\inted
  +\Li\left(\frac{2\sqrt\xi}{a(1+w)}\right)
  -\Li\left(\frac{2\sqrt\xi}{a(1-w)}\right)\,+\nonumber\\&&
  +\Li\left(\frac{1+w}2\right)-\Li\left(\frac{1-w}2\right)
  +\Li\left(\frac{a(1+w)}4\right)-\Li\left(\frac{a(1-w)}4\right)\\
t_6&:=&\ln^2(1+w)+\ln^2(1-w)+\ln\left(\frac{2+\sqrt\xi}8\right)\ln(1-w^2)
  \,+\nonumber\\&&
  +\Li\left(\frac{2\sqrt\xi}{a(1+w)}\right)
  +\Li\left(\frac{2\sqrt\xi}{a(1-w)}\right)
  -2\Li\left(\frac{2\sqrt\xi}a\right)\,+\nonumber\\&&
  +\Li\left(\frac{1+w}2\right)+\Li\left(\frac{1-w}2\right)
  -2\Li\left(\frac12\right)\,+\nonumber\\&&
  +\Li\left(\frac{a(1+w)}4\right)+\Li\left(\frac{a(1-w)}4\right)
  -2\Li\left(\frac a4\right)\\
t_7&:=&2\ln\left(\frac{1-\xi}{2\xi}\right)\inted
  -\Li\left(\frac{2v}{(1+v)^2}\right)
  +\Li\left(-\frac{2v}{(1-v)^2}\right)\,+\nonumber\\&&
  -\frac12\Li\left(-\left(\frac{1+v}{1-v}\right)^2\right)
  +\frac12\Li\left(-\left(\frac{1-v}{1+v}\right)^2\right)\,+\\&&
  +\Li\left(\frac{2w}{1+w}\right)-\Li\left(-\frac{2w}{1-w}\right)
  -2\Li\left(\frac{w}{1+w}\right)+2\Li\left(-\frac{w}{1-w}\right)
  \,+\nonumber\\&&
  +\Li\left(\frac{2aw}{b+aw}\right)-\Li\left(-\frac{2aw}{b-aw}\right)
  -2\Li\left(\frac{aw}{b+aw}\right)+2\Li\left(-\frac{aw}{b-aw}\right)
  \nonumber\\
t_8&:=&\ln\left(\frac\xi 4\right)\inted
  +\Li\left(\frac{2v}{1+v}\right)-\Li\left(-\frac{2v}{1-v}\right)-\pi^2\\
t_9&:=&2\ln\left(\frac{2(1-\xi)}{\sqrt\xi}\right)\inted
  +2\left(\Li\left(\frac{1+v}2\right)-\Li\left(\frac{1-v}2\right)\right)
  \,+\nonumber\\&&
  +3\left(\Li\left(-\frac{2v}{1-v}\right)
  -\Li\left(\frac{2v}{1+v}\right)\right)\\
t_{10}&:=&\ln\left(\frac4\xi\right),\quad
t_{11}\ :=\ \ln\left(\frac{4(1-\sqrt\xi)^2}\xi\right),\quad
t_{12}\ :=\ \ln\left(\frac{4(1-\xi)}\xi\right)
\eea

\vspace{2cm}
\centerline{\Large\bf Table Captions}
\vspace{.5cm}
\begin{list}{\bf\rm Tab.~1: }{
\labelwidth1.6cm \leftmargin2.5cm \labelsep0.4cm \itemsep0ex plus0.2ex }
\item Independent helicity components of the hadron tensor $H_{\mu\nu}$
in the spherical basis (column~2) and in the Cartesian basis (column~3).
Column~4 gives the respective angular coefficients that determine the
lepton-hadron correlations
\end{list}

\newpage 

\centerline{\Large\bf Figure Captions}
\vspace{.5cm}
\newcounter{fig}
\begin{list}{\bf\rm Fig.\ \arabic{fig}: }{\usecounter{fig}
\labelwidth1.6cm \leftmargin2.5cm \labelsep0.4cm \itemsep0ex plus0.2ex }
\item Definition of the polar angle $\theta$ and azimuthal angle $\chi$
decribing the relative orientation between lepton plane and hadron plane
\item Transverse polarization for the top-quark with mass $m_t=174$~GeV 
and running coupling constant $\as$, $\as(\MZ)=0.118$) at three different 
energies ($\sqrt{q^2}=360$~GeV (full line), 500~GeV (dashed) and 1000~GeV 
(dash-dotted)). Born term results drawn as dotted lines\\
(a) perpendicular polarization\quad (b) normal polarization
\item Transverse polarization for the bottom-quark on the $Z$-peak (with
bottom mass set to $m_b=4.83$~GeV \cite{trans7} and $\as(\MZ)=0.118$).\\
Full line: $O(\as^0)+O(\as)$; dotted line: Born term result\\
(a) perpendicular polarization\quad (b) normal polarization
\end{list}

\begin{table}[h]
\begin{center}
\begin{tabular}{|l||l|l|l|}
\hline
&spherical components&Cartesian components&angular factors\\
\hline\hline
$H_U$&$H_{++}+H_{--}$&$H_{11}+H_{22}$&$\frac38(1+\cos^2\theta)$\\
\hline
$H_L$&$H_{00}$&$H_{33}$&$\frac34\sin^2\theta$\\
\hline
$H_T$&$\frac12(H_{+-}+H_{-+})$&$\frac12(-H_{11}+H_{22})$
  &$\frac34\sin^2\theta\cos 2\chi$\\
\hline
$H_I$&$\frac14(H_{+0}+H_{0+}-H_{-0}-H_{0-})$
  &$\frac{-1}{2\sqrt 2}(H_{31}+H_{13})$
    &$\frac{-3}{2\sqrt 2}\sin 2\theta\cos\chi$\\
\hline
$H_9$&$-\frac i4(H_{+0}-H_{0+}-H_{-0}+H_{0-})$
  &$\frac{-i}{2\sqrt 2}(H_{31}-H_{13})$
    &$\frac3{\sqrt 2}\sin\theta\sin\chi$\\
\hline
$H_F$&$H_{++}-H_{--}$&$-i(H_{12}-H_{21})$&$\frac34\cos\theta$\\
\hline
$H_A$&$\frac14(H_{+0}+H_{0+}+H_{-0}+H_{0-})$
  &$\frac{-i}{2\sqrt 2}(H_{23}-H_{32})$
    &$\frac{-3}{\sqrt 2}\sin\theta\cos\chi$\\
\hline
$H_4$&$-\frac i2(H_{+-}-H_{-+})$&$-\frac12(H_{12}+H_{21})$
  &$-\frac34\sin^2\theta\sin 2\chi$\\
\hline
$H_5$&$-\frac i4(H_{+0}-H_{0+}+H_{-0}-H_{0-})$
  &$\frac{-1}{2\sqrt 2}(H_{23}+H_{32})$
    &$\frac3{2\sqrt 2}\sin 2\theta\sin\chi$\\
\hline
\end{tabular}
\end{center}
\bigskip
\centerline{\Large\bf Table 1}
\end{table}
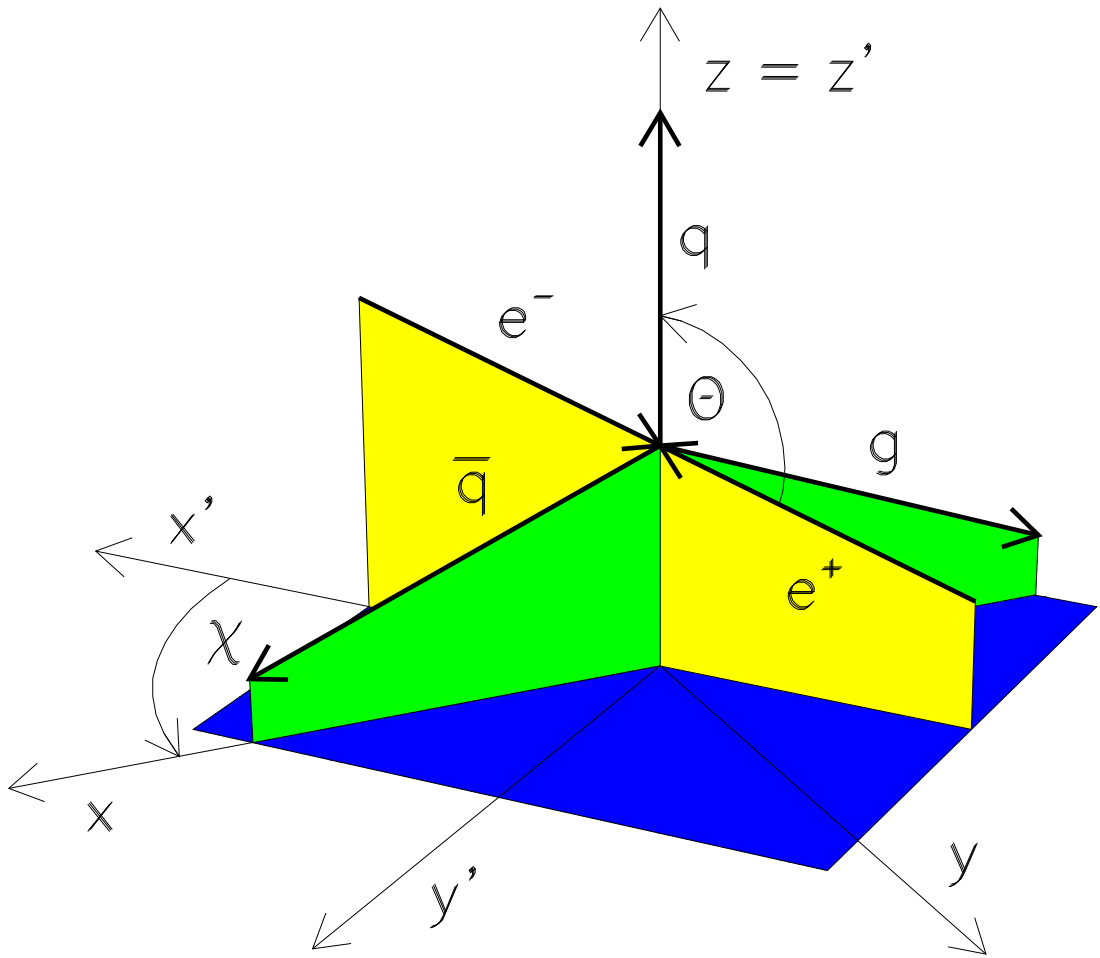
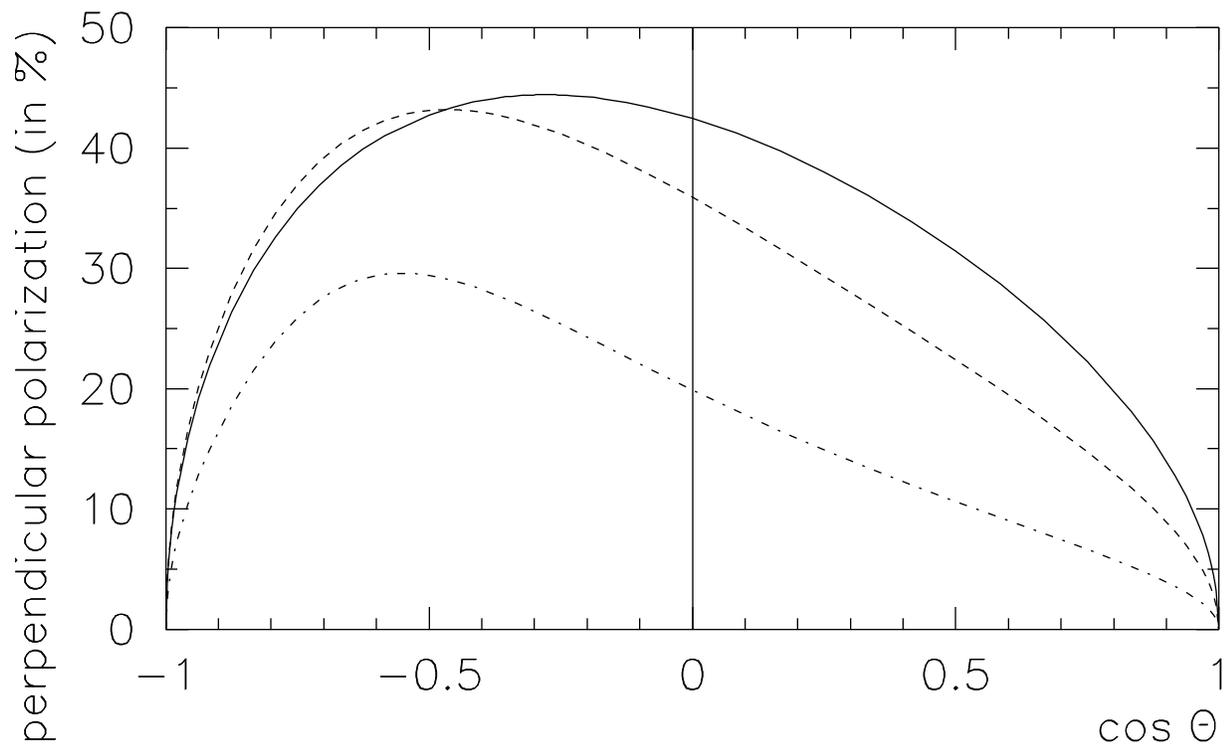
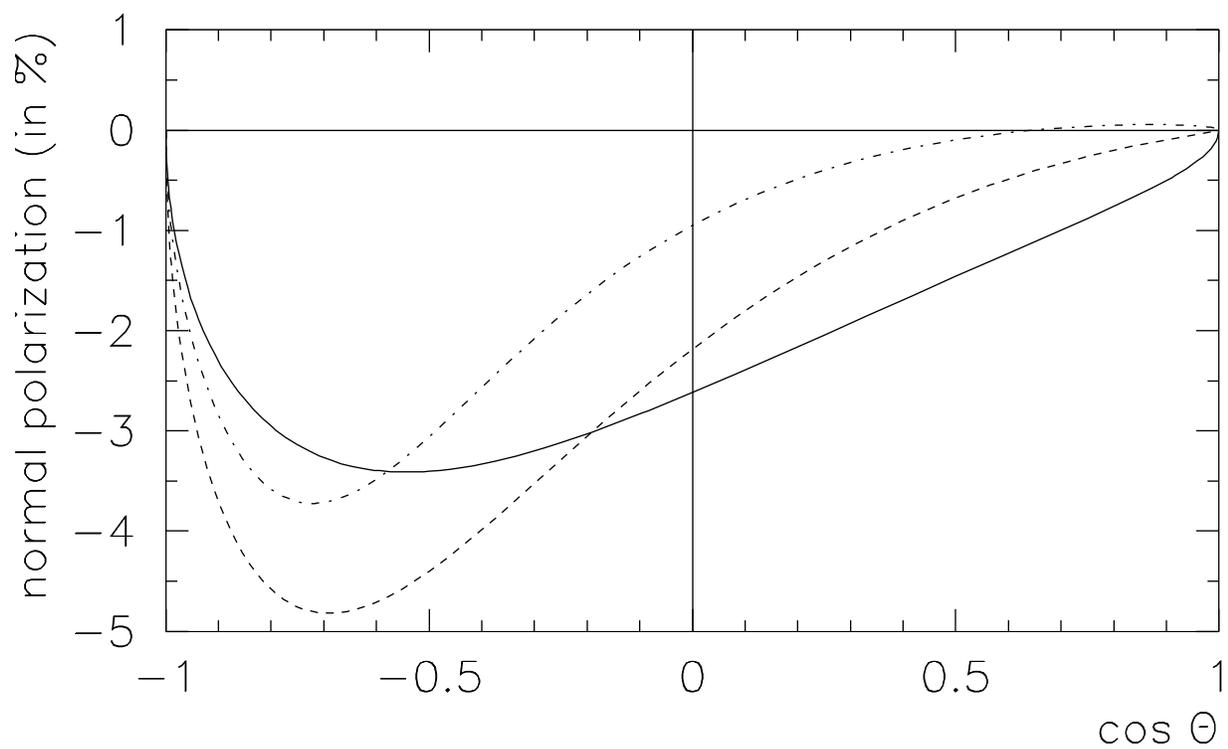
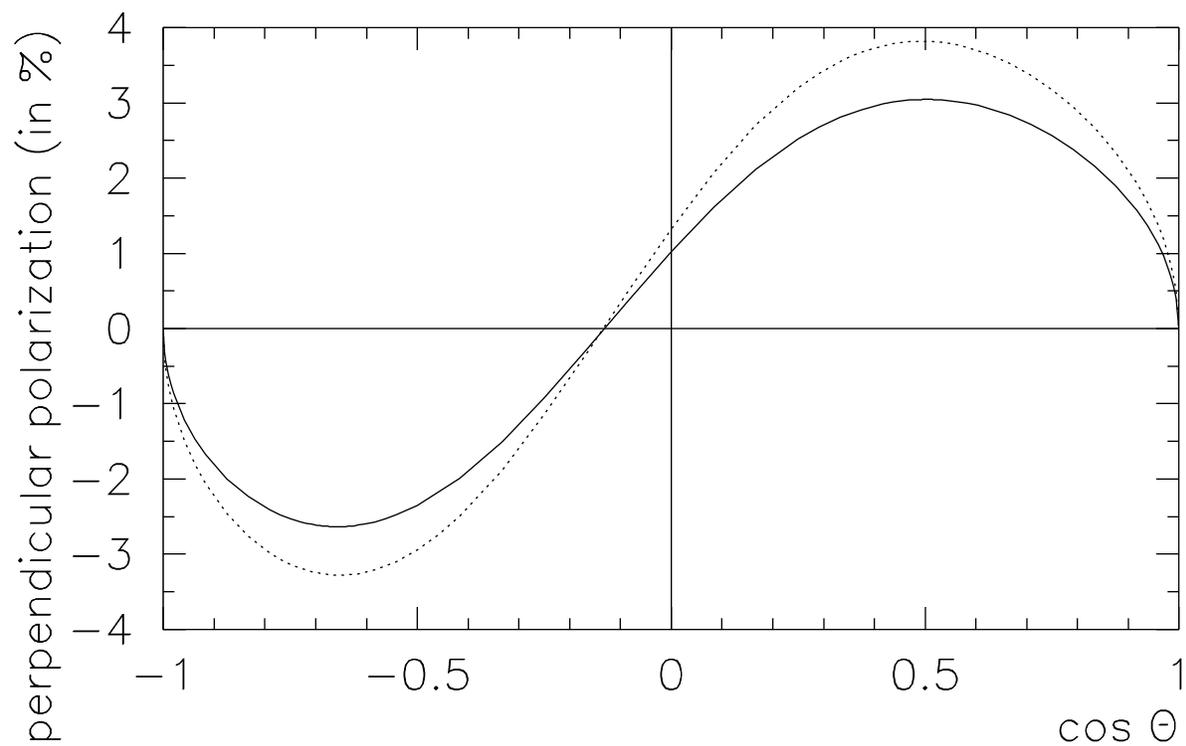
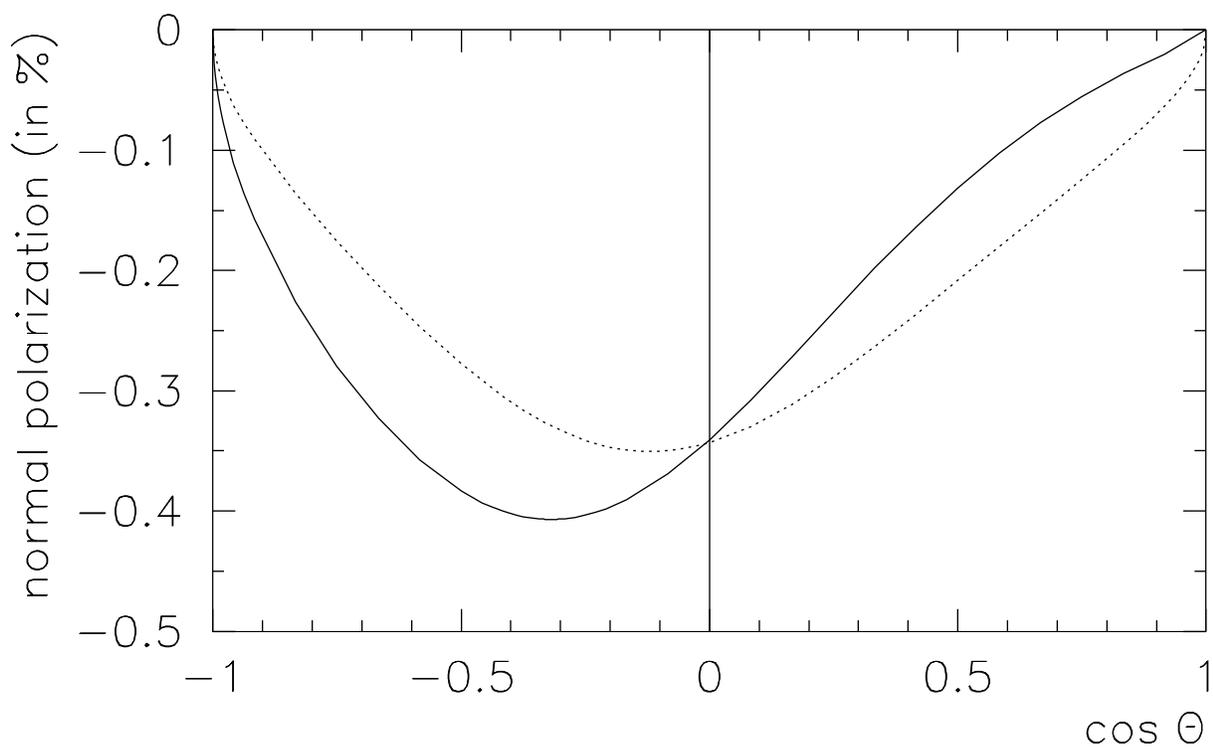
\end{document}